\documentclass[aps,prd,preprintnumbers,superscriptaddress,nofootinbib]{revtex4}
\usepackage[dvips]{graphicx}
\usepackage{bm,latexsym,amsmath,amssymb,amsfonts,mathrsfs}
\usepackage{color}
\input{colordvi.tex}
\newcommand*{\D}{{\rm d}}
\newcommand*{\mpl}{M_{\rm Pl}}
\begin{document}

\title{Black hole perturbation in the most general scalar-tensor theory
with second-order field equations II: the even-parity sector}

\author{Tsutomu~Kobayashi}
\affiliation{Department of Physics, Rikkyo University,
Toshima, Tokyo 175-8501, Japan
}

\author{Hayato~Motohashi}
\affiliation{Kavli Institute for Cosmological Physics, The University of Chicago, 
Chicago, Illinois 60637, USA
}

\author{Teruaki~Suyama}
\affiliation{Research Center for the Early Universe (RESCEU),
Graduate School of Science,
The University of Tokyo, Tokyo 113-0033, Japan
}

\begin{abstract}
We perform a fully relativistic analysis of even-parity linear perturbations 
around a static and spherically symmetric solution in the most general scalar-tensor 
theory with second-order field equations.
This paper is a sequel to Kobayashi {\em et al.} [Phys.\ Rev.\ D {\bf 85}, 084025 (2012)], in which the linear perturbation 
analysis for the odd-parity modes is presented.
Expanding the Horndeski action to second order in perturbations and
eliminating auxiliary variables, we derive
the quadratic action for even-parity perturbations written solely in terms of
two dynamical variables. The two perturbations can be interpreted as
the gravitational and scalar waves.
Correspondingly, we obtain two conditions to evade ghosts and
two conditions for the absence of gradient instabilities.
Only one in each pair of conditions
yields a new stability criterion,
as the conditions derived from the stability of the gravitational-wave degree of freedom
coincide with those in the odd-parity sector.
Similarly, the propagation speed of one of the two modes
is the same as that for the odd-parity mode, while
the other differs in general from them.
Our result is applicable to all the theories of gravitation
with an extra single scalar degree of freedom
such as the Brans-Dicke theory, $f(R)$ models, and Galileon gravity.
\end{abstract}

\preprint{RUP-14-4, RESCEU-5/14}
\maketitle

\section{Introduction}
Operation of the second-generation gravitational-wave detectors
such as Advanced LIGO \cite{Harry:2010zz}, Advanced Virgo \cite{Accadia:2010aa}, 
and KAGRA \cite{Somiya:2011np} in the coming decade will bring
detection of gravitational waves~(GWs)
and open a new era of gravitational wave astronomy.
Since typical GWs come from strong gravity regions such as the
vicinity of a black hole, the detection of GWs enables us to
test general 
relativity (GR)
and alternative theories of gravity in the strong fields.
In order to distinguish GR from other theories of gravity,
theoretical understanding
of fundamental properties of modified gravity is crucial.

It is interesting to consider,
among many directions of extending GR, a class of 
scalar-tensor theories for which field equations of the scalar field
and the metric are
at most of second order.
This class is known to be described by the Horndeski theory in a generic way,
in terms of four arbitrary functions of 
the scalar field $\phi$ and its kinetic term $(\partial \phi)^2$~\cite{Horndeski:1974wa}. 
The Horndeski theory includes
well studied models such as the Brans-Dicke theory, $f(R)$ theories,
Galileon models as specific cases.  
In Ref.~\cite{Kobayashi:2012kh},
fully relativistic study of linear perturbations around 
static and spherically symmetric spacetime has been performed,
restricting the analysis to the odd-parity perturbations, in
the Horndeski theory\footnote{See
Refs.~\cite{DeFelice:2011ka,Motohashi:2011pw,Motohashi:2011ds}
for black hole perturbations in other classes of
modified gravity theories.}.
Since no particular theory is assumed in its formulation, 
the perturbation equation and the stability conditions derived in~\cite{Kobayashi:2012kh}
have versatile and wide applicability.
It can also help us understand what kinds of differences are expected in
the behavior of perturbations in general within the Horndeski theory.
In the odd-parity sector, the scalar field is not perturbed and
only the metric perturbations enter the game.
As in the case of GR, it has been found that there is only one dynamical variable corresponding
to the GW degree of freedom. From the second-order Lagrangian for that variable,
we have obtained the conditions for the absence of ghost and gradient instabilities,
which put constraints on the form of the four functions involved
in the Horndeski action.
Variation of the second-order Lagrangian yields a wave equation, giving
the generalization of the Regge-Wheeler equation.

The aim of this paper is to extend our previous study~\cite{Kobayashi:2012kh} to 
the even-parity perturbations.
Since the scalar field perturbation participates in the even-parity sector
in addition to the metric perturbations,
it is expected that linearized field equations reduce to coupled differential 
equations for two dynamical variables corresponding to the GW and the scalar-field
degrees of freedom.
We will explicitly demonstrate that this is indeed the case, starting from the
second-order Lagrangian and eliminating the auxiliary fields by using the constraint equations.
One can see that the resultant wave equations are of second order both in
time and radial coordinates, which is reasonable given the fact that the original
field equations are of second order in the Horndeski theory.
Our second-order Lagrangian
provides two conditions for the absence of ghosts, one for the GWs
and the other for the scalar wave.
The condition for the former mode
is the same as that for the odd-parity mode,
while stability of the latter imposes
a completely new condition and thus
places an additional restriction to the Horndeski theory.
We also find that the propagation speed of the GWs coincides with that in the odd-parity sector,
giving
a consistency relation that holds in any second-order scalar-tensor theories.
However, the propagation speed of the scalar wave is generically different from that of the GWs.
This yields yet another
condition to avoid gradient instabilities, restricting further the form of
the Horndeski action.
Combining the results presented in this paper with the ones for the odd-parity perturbations,
we can test the viability of a given modified theory of gravity with
a single scalar degree of freedom.

The paper is organized as follows.
In the next section, we introduce the Lagrangian of the theories in the Horndeski class.
In Sec.~\ref{sec-even-perturbation} we compute the second-order Lagrangian
of the even-parity perturbations and derive the stability conditions as well
as the propagation speeds.
In Sec.~\ref{sec-application}, we apply our results to some specific models.
Section~\ref{conclusion} is devoted to the conclusion.
Appendices~\ref{sec-background} and \ref{app-a} provide the explicit form of the background equations and coefficients for the second-order Lagrangian, respectively.

\section{Horndeski theory and spherically symmetric background}

We start with defining the theory and the background metric we use.
The Horndeski theory~\cite{Horndeski:1974wa}
is the most general scalar-tensor theory with second-order field equations.
This theory was rederived recently~\cite{Deffayet:2011gz} in the course of
generalizing the Galileons~\cite{Nicolis:2008in},
and the modern form given in~\cite{Deffayet:2011gz}
was shown to be equivalent to the original Horndeski theory~\cite{Kobayashi:2011nu}.
The Horndeski theory is described by the following four
Lagrangians:
\begin{eqnarray}
{\cal L}_2&=&K(\phi, X),
\\
{\cal L}_3&=&-G_3(\phi, X)\Box \phi,
\\
{\cal L}_4&=&G_4(\phi,X)R
+G_{4X}\left[(\Box\phi)^2-(\nabla_\mu\nabla_\nu\phi)^2\right],
\\
{\cal L}_5&=&G_5(\phi, X)G_{\mu\nu}\nabla^\mu\nabla^\nu\phi
-\frac{1}{6}G_{5X}\left[ (\Box\phi)^3
-3\Box\phi(\nabla_\mu\nabla_\nu\phi)^2
+2(\nabla_\mu\nabla_\nu\phi)^3 \right],
\end{eqnarray}
where $K$ and $G_i$ are arbitrary functions of $\phi$ and $X:=-(\partial\phi)^2/2$.
Here we used the notation $G_{iX}$ for $\partial G_i/\partial X$.
Our action is thus
\begin{eqnarray}
S=\sum_{i=2}^5\int\D^4x\sqrt{-g}{\cal L}_i. \label{action1}
\end{eqnarray}
By choosing the functions $K$ and $G_i$ appropriately,
one can express any second-order scalar-tensor theory
in terms of the Horndeski theory. Some examples are presented in Ref.~\cite{Kobayashi:2011nu}.

We consider a static and spherically symmetric background solution
of the Horndeski theory. The background metric can be written as
\begin{eqnarray}
\D s^2=-A(r)\D t^2+\frac{\D r^2}{B(r)}+C(r) r^2\left(\D\theta^2+\sin^2\theta\,\D \varphi^2\right).
\label{bgmetric}
\end{eqnarray}
The scalar field is also dependent only on the radial coordinate, $\phi=\phi(r)$,
and hence $X=-B(\phi')^2/2$, where the prime denotes differentiation with
respect to $r$.
Without loss of generality, we can set $C(r)=1$.
Nevertheless, we introduced $C(r)$ because it is convenient
to retain $C(r)$ when deriving the field equations from the variational principle.
We will therefore impose this condition after taking variations to get the background
field equations.
The background equations are summarized in Appendix A.

In the analysis of linear perturbations around the above background,
we will not specify a concrete form of the solution $A(r), B(r)$, and $\phi(r)$,
as it is not necessary for deriving generic stability conditions against
perturbations.

\section{Formulation of the even-parity Perturbations}
\label{sec-even-perturbation}
\subsection{Decomposition of the even-parity perturbations}
Given the background equations of motion, we can now derive the quadratic action for perturbations.
In the Regge-Wheeler formalism \cite{Regge:1957td}, 
the metric perturbations are decomposed into odd- and even-type perturbations according 
to their transformation properties under the two-dimensional rotation.
Furthermore, each perturbation can be decomposed into the sum of spherical harmonics.
Then, at linear order in perturbation equations, or equivalently at second order
in the action for the perturbations, on the static and spherically symmetric background, 
the perturbation variables having different $\ell,~m$, and parity
do not mix each other. This fact drastically simplifies our perturbation analysis.
The odd-parity perturbations in the Horndeski theory have been
investigated in Ref.~\cite{Kobayashi:2012kh}. This paper is the sequel
of Ref.~\cite{Kobayashi:2012kh}, and
we will now concentrate on the even-parity perturbations.

The even-parity metric perturbations can be written as \cite{Regge:1957td} 
\begin{eqnarray}
h_{tt}&=&A(r)\sum_{\ell, m}H_{0,\ell m}(t,r)Y_{\ell m}(\theta,\varphi), \label{htt}\\
h_{tr}&=&\sum_{\ell, m}H_{1,\ell m}(t,r)Y_{\ell m}(\theta,\varphi),\\
h_{rr}&=&\frac{1}{B(r)}\sum_{\ell, m}H_{2,\ell m}(t,r)Y_{\ell m}(\theta,\varphi),\\
h_{ta}&=&\sum_{\ell, m}\beta_{\ell m}(t,r)\partial_{a}Y_{\ell m}(\theta,\varphi),\label{hta} \\
h_{ra}&=&\sum_{\ell, m}\alpha_{\ell m}(t,r)\partial_{a}Y_{\ell m}(\theta,\varphi),\label{hra} \\ 
h_{ab}&=&\sum_{\ell, m} K_{\ell m}(t,r) g_{ab} Y_{\ell m}(\theta,\varphi)+\sum_{\ell, m} G_{\ell m}(t,r) \nabla_a \nabla_b Y_{\ell m}(\theta,\varphi)\,. \label{hab}
\end{eqnarray}
The scalar field $\phi$ also has an even-parity perturbation,
\begin{equation}
\phi(t,r,\theta,\varphi)=\phi(r)+\sum_{\ell, m}\delta \phi_{\ell m}(t,r)Y_{\ell m}(\theta,\varphi).
\end{equation}
 
Because of the general covariance, we have gauge degrees of freedom,
enabling us to set some of the perturbation variables to zero by performing
an infinitesimal coordinate transformation $x^{\mu}\to x^{\mu}+\xi^{\mu}$.
Out of totally four degrees of freedom of the gauge transformation, 
three belong to the even-parity sector.
The three gauge functions can be written as \cite{Regge:1957td}
\begin{equation}
\xi_t=\sum_{\ell, m} T_{\ell m}(t,r)Y_{\ell m}(\theta,\varphi),~~~\xi_r=\sum_{\ell, m} R_{\ell m}(t,r)Y_{\ell m}(\theta,\varphi),~~~\xi_a=\sum_{\ell, m} \Theta_{\ell m}(t,r) \partial_a Y_{\ell m}(\theta,\varphi),
\end{equation}
where $T_{\ell m}(t,r),~R_{\ell m}(t,r)$, and $\Theta_{\ell m}(t,r)$ are arbitrary functions
of $t$ and $r$.
The transformation rule for each 
metric component under these gauge transformations
is given by~\cite{Regge:1957td}
\begin{eqnarray}
H_{0,\ell m}(t,r) &\to& H_{0,\ell m}(t,r)+\frac{2}{A} {\dot T_{\ell m}}(t,r)-\frac{A'B}{A} R_{\ell m}(t,r), \\
H_{1,\ell m}(t,r) &\to& H_{1,\ell m}(t,r)+{\dot R_{\ell m}}(t,r)+T_{\ell m}'(t,r)-\frac{A'}{A}T_{\ell m}(t,r), \\
H_{2,\ell m}(t,r) &\to& H_{2,\ell m}(t,r)+2BR_{\ell m}'(t,r)+B'R_{\ell m}(t,r), \\
\beta_{\ell m}(t,r) &\to& \beta_{\ell m}(t,r)+T_{\ell m}(t,r)+{\dot \Theta_{\ell m}}(t,r), \\
\alpha_{\ell m}(t,r) &\to& \alpha_{\ell m}(t,r)+R_{\ell m}(t,r)+\Theta_{\ell m}'(t,r)-\frac{2}{r} \Theta_{\ell m}(t,r), \\
K_{\ell m}(t,r) &\to& K_{\ell m}(t,r)+\frac{2B}{r} R_{\ell m}(t,r), \\
G_{\ell m}(t,r) &\to& G_{\ell m}(t,r)+2 \Theta_{\ell m}(t,r).
\end{eqnarray}
From these transformation rules we see that $G_{\ell m},~K_{\ell m}$,
and $\beta_{\ell m}$ can be set to zero by solving the coupled algebraic equations for
$\Theta_{\ell m},~R_{\ell m}$, and $T_{\ell m}$.
The solution is unique,
and hence the condition
$G_{\ell m}=K_{\ell m}=\beta_{\ell m}=0$ completely fixes the gauge.
In the following, we will use this gauge condition for
the calculation of the second-order action.

Note that the above argument about the gauge fixing does not apply to
the monopole ($\ell =0$)
and dipole ($\ell=1$) perturbations.
For the monopole perturbations, we have $\alpha=\beta=G=0$ identically and the
gauge transformation given by $\Theta$ is irrelevant.
For the dipole perturbations, $K$ and $G$ appear in $h_{ab}$ only through the combination
$K-G$, and hence the decomposition of $h_{ab}$ into the two components is in fact redundant.
The analysis for these two cases will be presented separately after studying
higher multipoles with $\ell \ge 2$.

\subsection{Brief review of the odd-parity perturbations}
Before going into the detailed investigation of the even-parity perturbations,
let us take a quick look at the main result of the analysis for the odd-parity perturbations
obtained in \cite{Kobayashi:2012kh}
in order to facilitate 
the comparison between the odd-parity and even-parity results.
The scalar field $\phi$ does not acquire odd-parity perturbations
and only the metric is perturbed.
After some manipulations of the second-order Lagrangian, 
it is confirmed that there is only one dynamical variable left in the final Lagrangian.
Varying this final Lagrangian, the master equation
for the odd-parity perturbations can be derived.
All the other perturbation variables are determined once the solution of the
master equation is obtained.
It is easy to check that the resultant master equation reduces to the Regge-Wheeler 
equation in the GR limit.
To evade ghost and gradient instabilities
it is required that
all the following conditions must be satisfied simultaneously:
\begin{eqnarray}
{\cal F} &:=& 2 \left( G_4+\frac{1}{2} B\phi' X' G_{5X}-X G_{5\phi} \right)>0, \label{con-F}\\
{\cal G} &:=& 2 \bigg[ G_4-2X G_{4X}+X \left( \frac{A'}{2A}
B\phi' G_{5X}+G_{5\phi} \right) \bigg]>0, \label{con-G}\\
{\cal H} &:=&  2 \bigg[ G_4-2X G_{4X}+X
\left( \frac{B \phi'}{r} G_{5X}+G_{5\phi} \right) \bigg]
=-\frac{2A}{B} \frac{\partial {\cal E}_C}{\partial A''} >0. \label{con-H}
\end{eqnarray}
Here, ${\cal E}_C$ is the quantity
introduced as the ``left-hand side'' of
the background equation and is defined in Appendix A.
The propagation speed along the radial direction is given by 
\begin{equation}
c_{\rm odd}^2=\frac{\cal G}{\cal F}.
\end{equation}
Since $\phi$ is not perturbed in the odd-parity sector,
$c_{\rm odd}$ can be interpreted as the propagation speed
of the GWs.
This fact can also be understood by noting that
$c_{\rm odd}$ depends only on $G_4$ and $G_5$, the functions
coupled to the curvature in the Horndeski action.

\subsection{Even-parity perturbations with $\ell \ge 2$}
Substituting into the action~(\ref{action1}) both the metric and the
scalar-field perturbations 
with the gauge choice $G_{\ell m}=K_{\ell m}=\beta_{\ell m}=0$,
we find that the second-order Lagrangian is given by
\begin{eqnarray}
\frac{2\ell+1}{2\pi} {\cal L}&=&H_0 \left[ a_1 \delta \phi''+a_2 \delta \phi'+a_3 H_2'+j^2 a_4 \alpha'+\left( a_5+j^2 a_6 \right) \delta \phi+\left( a_7+j^2 a_8 \right)H_2+j^2a_9 \alpha \right] \nonumber \\
&&+j^2 b_1 H_1^2+H_1 (b_2 {\dot {\delta \phi}}'+b_3 {\dot {\delta \phi}}+b_4 {\dot H_2}+j^2b_5 {\dot \alpha}) \nonumber \\
&&+c_1 {\dot H_2} {\dot {\delta \phi}}+H_2 \left[ c_2 \delta \phi'+\left( c_3+j^2 c_4 \right) \delta \phi+j^2 c_5 \alpha \right]+c_6 H_2^2+j^2d_1 {\dot \alpha}^2+j^2 \alpha (d_2 \delta \phi'+d_3 \delta \phi)+j^2 d_4 \alpha^2 \nonumber \\
&&+e_1 {\dot {\delta \phi}}^2+e_2 \delta \phi'^2+\left( e_3+j^2 e_4 \right) \delta \phi^2, \label{2nd-lag0}
\end{eqnarray}
where $j^2 := \ell (\ell+1)$.
Since only the perturbations in one multipole $(\ell,\,m)$ are considered at one time,
the suffixes $\ell,\,m$ of the perturbation variables are omitted without confusion. 
The explicit expression for the background dependent expansion coefficients 
$a_1,~a_2,~\cdots$ are presented in Appendix B.

The Lagrangian (\ref{2nd-lag0}) shows that both $H_0$ and $H_1$ are auxiliary fields.
In particular, no quadratic term in $H_0$ is present, and hence $H_0$
is a Lagrange multiplier,
giving rise to a constraint among the other
three variables, $H_2,~\alpha$, and $\delta \phi$:
\begin{equation}
a_1 \delta \phi''+a_2 \delta \phi'+a_3 H_2'+j^2 a_4 \alpha'+\left( a_5+j^2 a_6 \right) \delta \phi+\left( a_7+j^2 a_8 \right)H_2+j^2a_9 \alpha=0. \label{H0-constraint}
\end{equation}
Since $r$ derivatives of all the three variables appear in the above constraint,
this equation in its original form
cannot be solved for any one of $H_2$, $\alpha$, and $\delta\phi$.
In order to resolve this issue, we need to perform a
field redefinition and use a new variable $\psi$
defined by
\begin{equation}
H_2 = \frac{1}{a_3} \left( \psi-a_1 \delta \phi'-j^2 a_4 \alpha \right), \label{eq-H2}
\end{equation}
instead of $H_2$.
In terms of $\psi$, the first derivative of $\alpha$ as well as the
second derivative of $\delta \phi$ can be removed
simultaneously from Eq.~(\ref{H0-constraint}).
Thus, the constraint (\ref{H0-constraint}) becomes an algebraic equation for $\alpha$,
which can be solved to give
\begin{equation}
\alpha=\frac{1}{ j^2 a_4 [j^2 a_8 + (\frac{A'}{2 A} - \frac{1}{r}) a_3 ] } \bigg[ a_3 \psi'+j^2 a_8 \psi 
+\{ a_3 (a_2-a_1')-j^2 a_1 a_8 \} \delta \phi'+a_3 (a_5+j^2 a_6) \delta \phi \bigg]. \label{eq-alpha}
\end{equation}
Because of the existence of the quadratic term in $H_1$, 
variation with respect to $H_1$ gives
an equation that can be solved for $H_1$, yielding
\begin{equation}
H_1=-\frac{1}{2j^2 b_1} {(b_2 \delta \phi'+b_3 \delta \phi+b_4 H_2+j^2 b_5 \alpha)}\dot{\left.\right.},\label{eq-h1}
\end{equation}
where it should be understood that $H_2$ and $\alpha$ appearing in the above equation 
are replaced by $\psi$ and $\delta \phi$ using Eqs.~(\ref{eq-H2}) and (\ref{eq-alpha}).
Putting Eqs.~(\ref{eq-H2}), (\ref{eq-alpha}), and (\ref{eq-h1}) back
into Eq.~(\ref{2nd-lag0}) gives the reduced Lagrangian that
depends only on $\psi$ and $\delta \phi$.
At this stage, the resultant Lagrangian contains several higher derivative terms such as
${\dot {\delta \phi}} {\dot \psi}'$.
However, as it should be from the second-order nature of the Horndeski theory,
all such unwanted terms can be removed by performing some integration by parts.
As a result, we end up with the following Lagrangian containing at most first derivatives
and no mixing terms between $t$ and $r$ derivatives:
\begin{equation}
\frac{2\ell+1}{2\pi} {\cal L}=\frac{1}{2} {\cal K}_{ij} {\dot v}^i {\dot v}^j-\frac{1}{2} {\cal G}_{ij} {v^i}' {v^j}'-Q_{ij} v^i {v^j}'-\frac{1}{2} {\cal M}_{ij} v^i v^j, \label{fin-2nd-lagrangian}
\end{equation}
where $i$ and $j$ run from $1$ to $2$ and $v^1 := \psi,\;v^2 := \delta \phi$.

Let us first explore the conditions for the absence of ghost instabilities.
Since we have the two dynamical variables, the stability conditions
we are looking for are derived from
\begin{equation}
{\cal K}_{11}>0,\quad\det ({\cal K}) > 0.
\end{equation}
Explicitly, the first condition reads
\begin{equation}
{\cal K}_{11}=\frac{8 \sqrt{AB} {\left( 2r{\cal H}+\Xi \phi' \right)}^2}{\ell (\ell+1) A^2 {\cal H}^2} \frac{\ell (\ell+1){\cal P}_1-{\cal F}}{ {\left( 2r {\cal H} \ell(\ell+1)+{\cal P}_2 \right)}^2}>0, \label{k11}
\end{equation}
where ${\cal P}_1$ and ${\cal P}_2$ are defined respectively as
\begin{eqnarray}
{\cal P}_1&:=& \frac{B(2r{\cal H}+\Xi \phi')}{2A r^2 {\cal H}^2} {\left[ \frac{Ar^4 {\cal H}^4}{{\left( 2r{\cal H}+\Xi \phi' \right)}^2B} \right]}', \\
{\cal P}_2&:=&-B\left( 2-\frac{rA'}{A} \right) \left( 2r {\cal H}+\Xi \phi' \right),
\end{eqnarray}
and $\Xi$ as
\begin{eqnarray}
\Xi &:=&-\frac{r^2}{B} \frac{\partial{\cal E}_A}{\partial \phi''}=-\frac{2r^2A}{B}
\frac{\partial {\cal E}_\phi}{\partial A''} \nonumber \\ 
&=& 2r^2\left[ -XG_{3X}+\frac{2B\phi'}{r}\left\{ G_{4XY} - (XG_{5\phi})_X \right\} +  G_{4\phi Y} - \frac{1}{r^2}XG_{5X} +\frac{B}{r^2} (XG_{5X})_Y \right].\label{defXi}
\end{eqnarray}
In the definition of $\Xi$ we used again the ``left-hand side'' of the
background equations introduced in Appendix A.
We also introduced a notation
$f_Y := -2\sqrt{-X}\partial(\sqrt{-X} f)/\partial X$.
For instance, $f_Y=f+2Xf_X$ and $(Xf)_Y=X(3f+2Xf_X)$.
Notice that the $Y$-derivative does not commute with the $X$-derivative,
{\em i.e.,} $G_{4XY}\neq G_{4YX}$.
In deriving Eq.~\eqref{defXi} and
other quantities which we shall define later, one can use the following useful relations:
\begin{equation}
\frac{\partial [ (X^p f)']}{\partial \phi''}= -B \phi' (X^p f)_X, \quad \quad
\frac{\partial [ (\phi' X^p f)']}{\partial \phi''} = (X^p f)_Y.
\end{equation}

The second stability condition, ${\rm det}({\cal K})>0$, can be written explicitly as
\begin{equation}
\det ({\cal K})=\frac{16(\ell-1) (\ell+2){\left( 2r{\cal H}+\Xi \phi' \right)}^2 {\cal F} (2{\cal P}_1-{\cal F})}{\ell (\ell+1) A^2 {\cal H}^2 \phi'^2 {\left( 2r {\cal H} \ell(\ell+1)+{\cal P}_2 \right)}^2} > 0. \label{detk}
\end{equation}
With the first one of the stability conditions
for the odd-parity perturbations (\ref{con-H}), ${\cal F}>0$,
it is found that $\det ({\cal K})$ is positive if and only if 
\begin{equation}
2{\cal P}_1-{\cal F}>0. \label{even-no-ghost}
\end{equation}
It is interesting to note that if
Eq.~(\ref{even-no-ghost}) is satisfied then
Eq.~(\ref{k11}) is satisfied automatically, given that $\ell\ge 2$ and ${\cal F}>0$.
Consequently, only Eq.~(\ref{even-no-ghost}) gives rise to a
new independent condition for the absence of ghosts.

The squared propagation speeds of two modes along the radial direction,
$c_{s1}^2$ and $c_{s2}^2$,
are derived from the eigenvalues of the matrix ${(AB)}^{-1} {\cal K}^{-1} {\cal G}$
and are given by
\begin{eqnarray}
c_{s1}^2&=&\frac{\cal G}{\cal F}, \label{speed-c1} \\
c_{s2}^2&=&\frac{2r^2\Gamma {\cal H} \Xi \phi'^2-{\cal G} \Xi^2 \phi'^2- 4r^4
\Sigma {\cal H}^2/B}{{(2r {\cal H}+\Xi \phi')}^2 (2 {\cal P}_1-{\cal F})}, \label{speed-c2}
\end{eqnarray}
where $\Gamma$ and $\Sigma$ are defined as
\begin{eqnarray}
\Gamma&:=&\Gamma_1+\frac{A'}{A}\Gamma_2=-\frac{2}{B} \frac{\partial {\cal E}_C}{\partial \phi''}, \\
\Gamma_1&:=&
4\left[
-XG_{3X} + G_{4\phi Y} + \frac{B\phi'}{r} \left\{ G_{4XY} - (X G_{5\phi})_X \right\} \right] , \\
\Gamma_2&:=&
2B\phi'\left[
G_{4XY} - (XG_{5\phi})_X - \frac{B\phi'}{2rX} (XG_{5X})_Y
\right] 
=\frac{\partial {\cal H}'}{\partial \phi''}, \\
\Sigma&:=& \frac{X}{B} \frac{\partial {\cal E}_\phi}{\partial \phi''} \nonumber \\
&=& X\bigg[ \bigg.
K_{XY} - B\phi' \left(\frac{4}{r}+\frac{A'}{A}\right) (XG_{3X})_X - 2 G_{3\phi Y}
+2 \left(\frac{1-B}{r^2}-\frac{B}{r}\frac{A'}{A}\right) G_{4XY}
-\frac{4B}{r}\left(\frac{1}{r}+\frac{A'}{A}\right) (XG_{4XX})_Y
\nonumber\\
&&
+2B\phi' \left(\frac{4}{r}+\frac{A'}{A}\right)(XG_{4\phi X})_X
-\frac{B\phi'(1-3B)}{r^2}\frac{A'}{A}(XG_{5X})_X 
+\frac{2B^2\phi'}{r^2}\frac{A'}{A} (X^2G_{5XX})_X
\nonumber\\
&&
-2\left(\frac{1-B}{r^2}-\frac{B}{r}\frac{A'}{A}\right)G_{5\phi Y}
+\frac{2B}{r}\left(\frac{1}{r}+\frac{A'}{A}\right)(XG_{5\phi X})_Y 
\nonumber\\
&&
+2XG_{3\phi X}+B\phi'\left(\frac{4}{r}+\frac{A'}{A}\right)G_{4\phi X}+\frac{2}{r^2}XG_{5\phi X}
\bigg. \bigg].
\end{eqnarray}
The stability conditions for
the odd-parity modes, Eqs.~(\ref{con-F}) and (\ref{con-G}),
ensure that $c_{s1}^2>0$.
Using the no-ghost condition, Eq.~(\ref{even-no-ghost}), we see that
$c_{s2}^2$ is positive if and only if the
following condition is satisfied:
\begin{equation}
2r^2\Gamma {\cal H} \Xi \phi'^2-{\cal G} \Xi^2 \phi'^2-\frac{4r^4}{B} \Sigma {\cal H}^2>0.
\end{equation}
Since $c_{s1}$ depends only on the two of the Horndeski functions,
$G_4$ and $G_5$, and it is those two functions that are coupled to
the curvature in the action, $c_{s1}$ can be interpreted as the
propagation speed of GWs.
On the other hand, $c_{s2}$ involves both $K$ and $G_3$ as well,
and hence it is reasonable to
interpret $c_{s2}$ as the propagation speed of a scalar wave.
These interpretations are also supported by
the propagation speeds of the monopole and
dipole perturbations which will be computed shortly:
monopole and dipole modes arise entirely due to
the scalar degree of freedom,
and it will turn out that the two modes indeed propagate at $c_{s2}$.

Interestingly, $c_{s1}$ exactly coincides with $c_{\rm odd}$,
namely, odd-type and even-type GWs propagate at the same speed (though
it is not necessarily equal to the speed of light).
If future experiments would reveal that the consistency relation,
$c_{\rm odd}=c_{s1}$, is violated,
all the modified gravity theories in the Horndeski class
as well as GR could be excluded.

The mass matrix ${\cal M}_{ij}$ and the antisymmetric matrix $Q_{ij}$
provide further conditions for the stability of
static and spherically symmetric solutions.
However, explicit expressions for ${\cal M}_{ij}$ and $Q_{ij}$ are found to be
too complicated to be illuminating, and
we have not been able to give sufficiently concise stability conditions
from those matrix elements.

\subsection{Monopole perturbation: $\ell=0$}
For the monopole perturbations, $\alpha,~\beta$, and $G$ identically vanish 
in Eqs.~(\ref{hta}), (\ref{hra}), and (\ref{hab}).
The gauge functions that are still meaningful are $\xi_t$ and $\xi_r$.
As is the case for higher multipoles with $\ell \ge 2$, 
$\xi_r$ is fixed completely by setting $K$ to zero.
As for $\xi_t$, it can in principle be used to eliminate either $H_0$ or $H_1$.
However, since that is not a complete gauge fixing, we defer it until we 
derive the perturbation equations from the second-order Lagrangian. 
Keeping this in mind, the second-order Lagrangian for the monopole perturbations 
can be obtained by setting $j^2=0$ in the Lagrangian~(\ref{2nd-lag0}) as
\begin{eqnarray}
\frac{2\ell+1}{2\pi} {\cal L}&=&H_0 \left(a_1\delta\phi'-\frac{A}{2}b_3\delta\phi-\frac{A}{2}b_4H_2\right)'+\frac{b_2}{a_1}H_1 \left(a_1\delta\phi'-\frac{A}{2}b_3\delta\phi-\frac{A}{2}b_4H_2\right) \dot{\Big.\Big.} \nonumber \\
&&+c_1 {\dot H_2} {\dot {\delta \phi}}+H_2 (c_2 \delta \phi'+c_3 \delta \phi)+c_5 H_2^2+e_1 {\dot {\delta \phi}}^2+e_2 \delta \phi'^2+e_3 \delta \phi^2, \label{Lagmono}
\end{eqnarray}
where we have used the background equations to rewrite the first term.
We see from Eq.~(\ref{Lagmono}) that
there are no terms quadratic in $H_0$ and $H_1$ for $\ell=0$,
and hence those two variables are Lagrange multipliers in this case.
As a result, we obtain two constraint equations.
However, as is clear from Eq.~(\ref{Lagmono}),
the two constraints are not independent
but merge into the following single constraint in the end:
\begin{equation}
H_2=-\frac{b_3}{b_4}\delta\phi+\frac{2}{Ab_4}(a_1\delta\phi'+C_0), \label{mono-H2}
\end{equation}
where the integration constant $C_0$ amounts to the shift of one of the integration constants 
in the background solution.
Since we are interested in the perturbations that do not correspond to a mere change 
of the background solution, we set $C_0=0$.
Substituting Eq.~(\ref{mono-H2}) back into Eq.~\eqref{Lagmono} and
performing integration by parts, we arrive at
\begin{equation}
\frac{2\ell+1}{2\pi} {\cal L}=\frac{1}{2}\mathcal{K}_0 \dot{\delta \phi}^2- \frac{1}{2}\mathcal{G}_0 \delta \phi'^2-\frac{1}{2}\mathcal{M}_0 \delta \phi^2,
\end{equation}
where
$\mathcal{K}_0$, $c_s^2:= (AB)^{-1}\mathcal{K}_0^{-1}\mathcal{G}_0$,
and $\mathcal{M}_0$ are defined as
\begin{eqnarray}
\mathcal{K}_0&=& \frac{4}{\sqrt{AB}\phi'^2} \left( 2{\cal P}_1-{\cal F} \right), \\
c_s^2&=&\frac{2r^2\Gamma {\cal H} \Xi \phi'^2-{\cal G} \Xi^2 \phi'^2-4r^4
\Sigma {\cal H}^2/B}{{(2r {\cal H}+\Xi \phi')}^2 (2 {\cal P}_1-{\cal F})}, \\
\mathcal{M}_0&=& -\frac{1}{a_3^2} \big[ 2a_2^2 a_3 c_6-a_2 \{ -4a_1 a_7 c_6+a_3^2 (2c_3-c_2')+a_3
(a_7c_2+6c_6a_1'+2a_1 c_6')\} \nonumber \\
&&+a_3 \{ 2a_3^2 e_3+a_1'(a_7c_2+4c_6a_1')+a_3 (3c_3a_1'-a_1'c_2'+c_2(a_2'-a_1''))\} \nonumber \\
&&+a_1 \{ -4a_7 c_6a_1'+a_3^2 c_3'+a_3 (-a_7c_3-2c_6a_2'+2a_1'c_6'+2c_6a_1'')\} \big].
\end{eqnarray}
The no-ghost condition is given by $2 {\cal P}_1-{\cal F}>0$, which is exactly the same
as the one for higher multipoles with $\ell \ge 2$.
The propagation speed also coincides with $c_{s2}$ given in Eq.~(\ref{speed-c2}).
Since only the scalar wave is excited as a monopole perturbation,
this result allows us to interpret $c_{s1}$ and $c_{s2}$ as the propagation
speeds of gravitational and scalar waves, respectively.

\subsection{Dipole perturbation: $\ell=1$}
In the $\ell =1$ case, it can be checked that the metric perturbations
$h_{ab}$ depend on $K$ and $G$
only through the combination $K-G$.
Therefore, we may set $K=0$ from the outset.
By using the gauge transformation $\Theta$, we can set $G=0$.
We can also set $\beta=0$ by invoking $T$.
We still have a freedom to choose $R$, which can be used to set $\delta \phi=0$, and
finally we are left with
the four variables, $H_0,~H_1,~H_2$,
and $\alpha$.
Thus, the second-order Lagrangian for $\ell =1$ is
\begin{eqnarray}
\frac{2\ell+1}{2\pi} {\cal L}&=&H_0 \left[ a_3 H_2'+2 a_4 \alpha'+\left( a_7+2 a_8 \right)H_2+2a_9 \alpha \right] +2 b_1 H_1^2+H_1 (b_4 {\dot H_2}+2b_5 {\dot \alpha}) \nonumber \\
&&+2c_5 H_2 \alpha +c_6 H_2^2+2d_1 {\dot \alpha}^2+2 d_4 \alpha^2. \label{Lag-dipole}
\end{eqnarray}
As in the case of $\ell \ge 2$, $H_0$ is a Lagrange multiplier
whose variation yields a constraint between $H_2$ and $\alpha$.
Since $r$ derivatives of both $H_2$ and $\alpha$ appear in the constraint equation,
we perform a field redefinition
\begin{equation}
H_2 = \frac{1}{a_3} \left( \psi-2 a_4 \alpha \right), \label{dipole-eq-H2}
\end{equation}
and use the new variable $\psi$ to remove the $r$ derivative terms.
Note that Eq.~(\ref{dipole-eq-H2}) is
deduced from Eq.~(\ref{eq-H2}) by setting $\delta \phi=0$.
Variation with respect to $H_1$ gives 
\begin{equation}
H_1=-\frac{1}{4b_1} (b_4 {\dot H_2}+2b_5 {\dot \alpha}).\label{dipole-eq-H1}
\end{equation}
Substituting Eqs.~(\ref{dipole-eq-H2}) and~(\ref{dipole-eq-H1})
into the Lagrangian (\ref{Lag-dipole}), we obtain
the final result written solely in terms of $\psi$:
\begin{equation}
\frac{2\ell+1}{2\pi} {\cal L}=\frac{1}{2} {\cal K}_1 {\dot \psi}^2-\frac{1}{2} {\cal G}_1 \psi'^2
-\frac{1}{2}{\cal M}_1 \psi^2,
\end{equation}
where ${\cal K}_1$ and ${\cal G}_1$ are given by
\begin{eqnarray}
{\cal K}_1 &=& \frac{4 \sqrt{AB} {\left( 2r{\cal H}+\Xi \phi' \right)}^2}{A^2 {\cal H}^2} 
\frac{2{\cal P}_1-{\cal F}}{ {\left( 4r {\cal H}+{\cal P}_2 \right)}^2}, \\
{\cal G}_1 &=& \frac{4B^{3/2}}{\sqrt{A} {\cal H}^2}
\frac{\left( 2r^2\Gamma {\cal H} \Xi \phi'^2-{\cal G} \Xi^2 \phi'^2-
4r^4 \Sigma {\cal H}^2 /B\right)}{ {\left( 4r {\cal H}+{\cal P}_2 \right)}^2}.
\end{eqnarray}
The no-ghost condition is given by
$2{\cal P}_1-{\cal F} >0$, which again coincides with the one
for the other multipoles.
The propagation speed along the radial direction,
$c_s^2:= (AB)^{-1}{{\cal K}_1}^{-1}{\cal G}_1$,
is also the same as $c_{s2}^2$.
This is consistent with the interpretation that $c_{s2}$ corresponds to the propagation
speed of a scalar wave.

\section{Application to specific models}
\label{sec-application}
\subsection{General relativity}
As a first example, let us consider the simplest case,
{\it i.e.}, GR without a scalar field $\phi$.
This case amounts to setting
$G_4=\mpl^2/2$ and $K=G_3=G_5=0$, leading to
${\cal F}={\cal G}={\cal H}=2{\cal P}_1=\mpl^2>0$. Note that $c^2_{\rm odd}=1$, which means that the odd mode propagates the speed of light, and $2{\cal P}_1-{\cal F}=0$,
which means that one of the even modes does not propagate, as expected.

The background metric is given by the Schwarzschild metric, $A(r)=B(r)=1-2M/r$.
Since we do not have the scalar degree of freedom in the case of GR, the number of
degrees of freedom is reduced by one from the general case.
Nonetheless, all the procedures to arrive at Eq.~(\ref{fin-2nd-lagrangian}) 
are well defined in the GR limit.
Therefore, Eq.~(\ref{fin-2nd-lagrangian}) is still valid in GR and reduces to
\begin{equation}
\frac{2\ell+1}{2\pi}{\cal L}= \frac{2(j^2-2)}{\mpl^2}  \left[
\frac{1}{A} {\dot \Psi}^2
-A \Psi'^2- \frac{3+j^2-j^4+j^6-3{(1+j^2)}^2A+9(1+j^2)A^2-9A^3}{r^2 {(j^2+1-3A)}^2} \Psi^2 \right],
\end{equation}
where we used a new field $\Psi$ defined by $\psi=j(1+j^2-3A)\Psi$.
Introducing the tortoise coordinate,
$r_*:=\int dr/A(r)$, we find that the equation of motion
for $\Psi$ is given by
\begin{equation}
\frac{\partial^2 \Psi}{\partial t^2}-\frac{\partial^2 \Psi}{\partial r_*^2}
-\frac{A}{r^2 {(j^2+1-3A)}^2} 
\left( 3+j^2-j^4+j^6-3{(1+j^2)}^2A+9(1+j^2)A^2-9A^3 \right) \Psi=0.
\end{equation}
The Zerilli equation~\cite{Zerilli:1970se} is thus reproduced.

\subsection{Nonminimally coupled scalar field}

Let us next consider models
in which the scalar field is nonminimally
coupled to the Ricci scalar,
corresponding to the following choice of the Horndeski functions:
\begin{equation}
K=X,~~~G_3=0,~~~G_4=f(\phi),~~~G_5=0.
\end{equation}
In this case, we have
\begin{equation}
{\cal F}={\cal G}={\cal H}=2f,~~~2{\cal P}_1-{\cal F}=
\frac{2r^2f (f+3f_\phi^2) \phi'^2}{{(2f+rf_\phi \phi')}^2}.
\end{equation}
Thus, it is sufficient to impose $f>0$ in order to avoid ghost instabilities
in the odd-parity and even-parity sectors.
This is consistent with the naive expectation
that the kinetic term for the graviton have the wrong sign
and hence will be plagued by ghosts for $f<0$.
Note that the condition $f>0$ depends on the profile of $\phi$ but
not on the concrete form of the metric.
As for the propagation speeds, we find
\begin{equation}
c_{s1}^2=c_{s2}^2=1.
\end{equation}
Thus, perturbations propagate at the speed of light.

\subsection{Bocharova-Bronnikov-Melnikov-Bekenstein solution}
To give an explicit example of a black hole solution with scalar hair,
we consider the theory with a conformally coupled scalar field:
\begin{equation}
S=\frac{\mpl^2}{2} \int d^4x~\sqrt{-g}R-\int d^4x~\sqrt{-g} \left( \frac{1}{2} \partial^\mu \phi \partial_\mu \phi
+\frac{R}{12}\phi^2 \right).
\end{equation}
This is the special case of the previous example:
\begin{equation}
K=X,~~~G_3=0,~~~G_4=f(\phi)
=\frac{\mpl^2}{2}-\frac{\phi^2}{12},~~~G_5=0.
\end{equation}
An exact black hole solution with a nontrivial scalar-field configuration
has been found by Bocharova, Bronnikov, and Melnikov~\cite{Bocharova:1970pw}
and independently by Bekenstein~\cite{Bekenstein:1974sf}. The solution is given by
\begin{eqnarray}
&&ds^2=-{\left( 1-\frac{M}{r} \right)}^2 \D t^2+
\frac{\D r^2}{{ \left( 1-M/r \right)}^2}+r^2 \D\Omega^2,\\
&&\phi=\pm \frac{\sqrt{6} \mpl M}{r-M},
\end{eqnarray}
where $M$ is a constant.
The metric is exactly the same as the extremal Reissner-Nordstr\"{o}m metric 
and the horizon is located at $r=M$.
Classical stability of the
Bocharova-Bronnikov-Melnikov-Bekenstein~(BBMB) solution against the monopole perturbation 
has been addressed in~\cite{Bronnikov:1978mx,McFadden:2004ni,Konoplya:2005et}.
For $\phi^2 >6\mpl^2$ we have $f<0$, giving rise to ghosts.
In terms of $r$, ghosts appear for $r <2M$.
Thus, the BBMB solution is quantum mechanically unstable for $r<2M$.
Note that this unstable region is outside the horizon, as the horizon is
at $r=M<2M$. Explicitly, we have
\begin{eqnarray}
{\cal F}={\cal G}={\cal H}=\frac{\mpl^2r(r-2M)}{(r-M)^2},
\quad
2{\cal P}_1-{\cal F}=\frac{3\mpl^2M^2r(r-2M)}{(r^2-3Mr+3M^2)^2}.
\end{eqnarray}

\subsection{Models with a trivial scalar field configuration}
In some classes of scalar-tensor theories, no-hair theorems for black holes
have been established under certain assumptions~\cite{Hawking:1972qk,Bekenstein:1995un,Sotiriou:2011dz,Hui:2012qt}.
In light of this, let us consider
models in which the trivial solution with $\phi=\phi_0={\rm const}$ exists.
Assuming that $K,~K_{\phi}, \cdots$ are not singular at $\phi'=0$,
it is verified that $K=K_{\phi}=0$ must be satisfied at $\phi=\phi_0$
in order for the asymptotically flat background solution
with $\phi=\phi_0$ everywhere to exist.
In such theories,
the background equations of motion (\ref{back-eom}) shows that $A$ and $B$ 
are uniquely determined as
\begin{equation}
A(r)=B(r)=1-\frac{\mu}{r},
\end{equation}
where $\mu$ is an integration constant.
Thus, the metric takes the form of Schwarzschild
and $r=\mu$ is the horizon location.
In the present case, ${\cal F}={\cal G}={\cal H}=2G_4$ and 
we obtain the coefficients of
the final second-order Lagrangian \eqref{fin-2nd-lagrangian} as\footnote{Although
Eq.~(\ref{detk}) contains $\phi'$ in the denominator,
this potentially dangerous term is canceled out by
$2{\cal P}_1-{\cal F}$ in the numerator and
the final result remains finite.
Since the mathematical procedure to derive Eq.~\eqref{fin-2nd-lagrangian}
involves division by $\phi'$, we need to repeat the derivation of
Eq.~(\ref{fin-2nd-lagrangian}) starting from
Eq.~(\ref{2nd-lag0}) by setting $\phi'=0$ from the outset.
All the results in this subsection are derived by such a manipulation.}
\begin{eqnarray}
{\cal K}_{11}&=&
\frac{2 \left(j^2-2\right) r^3}{G_4 j^2 (r-\mu) \left[3 \mu+\left(j^2-2\right) r\right]^2}, \\
\det ({\cal K})&=&\frac{4 \left(j^2-2\right) r^6 \left(K_X G_4-2 G_{3\phi} G_4+3 G_{4\phi}^2\right)}{G_4^2 j^2
   (r-\mu)^2 \left[3 \mu+\left(j^2-2\right) r\right]^2}.
\end{eqnarray}
We find that if $G_4>0$,
${\cal F}$, ${\cal G}$, and ${\cal H}$ are positive
and ${\cal K}_{11}$ is also
positive outside the horizon, $r>\mu$.
The only nontrivial no-ghost condition is obtained
from $\det ({\cal K})>0$:
\begin{equation}
K_X G_4-2 G_{3\phi} G_4+3 G_{4\phi}^2>0.
\end{equation}
The propagation speeds are given by
\begin{equation}
c_{s1}^2=c_{s2}^2=1,
\end{equation}
{\em i.e.,} the two modes propagate at the speed of light.

Note that $G_5$ does not appear at all in the second-order Lagrangian in this case.
If the background solution has no scalar hair, {\it i.e.} $\phi'=0$,
$G_5$ is irrelevant to the background configuration and still lurks at the level of
linear perturbations.

\subsection{Derivative coupling to the Einstein tensor}
\begin{figure}[tb]
  \begin{center}
    \includegraphics[width=90mm]{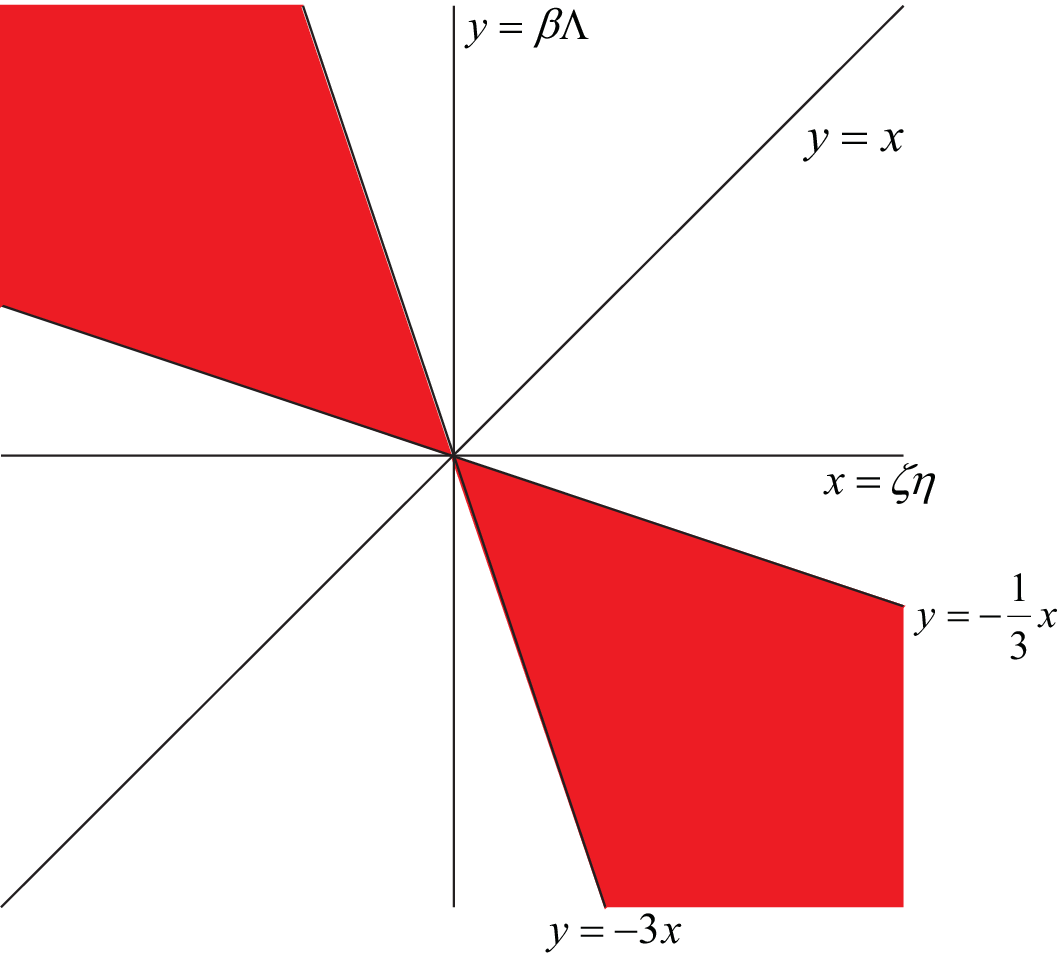}
  \end{center}
  \caption{The stable region is colored red. $\zeta >0$ is assumed and $\Lambda <0$ must be satisfied.}
  \label{fig:allow-region}
\end{figure}

Finally, we provide an example of the scalar-tensor theory with 
a derivative coupling to the Einstein tensor:
\begin{equation}
S=\int d^4x~\sqrt{-g}\left[ \zeta R-\eta \partial^\mu \phi \partial_\mu \phi +\beta G^{\mu\nu} \partial_\mu \phi \partial_\nu \phi
-2\Lambda \right].
\end{equation}
This action corresponds to the choice
\begin{equation}
K=2(\eta X-\Lambda),~~~G_3=0,~~~G_4=\zeta,~~~G_5=-\beta\phi.
\end{equation}
A static solution in this theory is given by~\cite{Babichev:2013cya,Rinaldi:2012vy,Anabalon:2013oea,Minamitsuji:2013ura}
\begin{eqnarray}
A&=&1-\frac{\mu}{r} +\frac{\eta}{3\beta} \frac{2\zeta \eta-\lambda}{2\zeta \eta+\lambda} 
r^2+\frac{\lambda^2}{4\zeta^2 \eta^2-\lambda^2} \frac{\arctan (r\sqrt{\eta /\beta})}{r\sqrt{\eta /\beta}}, \\
B&=&\frac{(\beta+\eta r^2)A}{\beta {(rA)}'}, \\
\phi'^2&=&-\frac{r \lambda {(r^2 A^2)}'}{2 {(\beta+\eta r^2)}^2 A^2},
\end{eqnarray}
where $\lambda =\zeta \eta+\beta \Lambda$ and $\mu$ is an integration constant.
For this solution, we find 
\begin{eqnarray}
{\cal F}&=&\frac{2\beta \zeta+(3\zeta \eta+\beta \Lambda) r^2}{\beta+\eta r^2}, \\
{\cal G}&=&{\cal H}=\frac{2\beta \zeta+(\zeta \eta-\beta \Lambda) r^2}{\beta+\eta r^2},
\end{eqnarray}
and therefore,
\begin{equation}
c_{s1}^2=\frac{2\beta \zeta+(\zeta \eta-\beta \Lambda) r^2}{2\beta \zeta+(3\zeta \eta+\beta \Lambda) r^2}. \label{Gkin-c1}
\end{equation}
Further conditions for stability can be obtained from
$2 {\cal P}_1-{\cal F}>0$ and $c_{s2}^2>0$, but
the full expressions are involved so that we only give
their large $r$ behavior:
\begin{eqnarray}
2 {\cal P}_1-{\cal F} &\xrightarrow{r\to \infty}& -\frac{4 {(\zeta \eta+\beta \Lambda)}^2}{\eta (\zeta \eta+3\beta \Lambda)}, \\
c_{s2}^2 &\xrightarrow{r\to \infty}& \frac{3 (-\zeta \eta+\beta \Lambda)}{\zeta \eta+3\beta \Lambda}. \label{Gkin-c2}
\end{eqnarray}
Let us impose that $t$ is a time-like coordinate at large $r$,
which puts the following condition: $A(r)>0$ at large $r$, namely,
\begin{equation}
\frac{\eta}{\beta} \frac{\zeta \eta-\beta \Lambda}{3\zeta \eta+\beta \Lambda}>0.
\end{equation}
Now, the stability conditions both for the odd-parity and even-parity perturbations
are summarized as
\begin{equation}
\frac{3\zeta \eta+\beta \Lambda}{\eta}>0,~~~\frac{\zeta \eta-\beta \Lambda}{\eta}>0,~~~
\eta (\zeta \eta+3\beta \Lambda)<0. \label{Gkin-stability}
\end{equation}
The stable parameter region is shown in Fig.~\ref{fig:allow-region}.

\begin{table}[htb]
  \caption{No-ghost conditions and propagation speeds for some specific models are summarized. }
  \label{table-1}
  \begin{tabular}{|c|c|c||c|} \hline
    Model & No-ghost conditions & Propagation speeds & Remarks \\ \hline \hline
    General Relativity & No ghost & Speed of light & Reduces to the Zerilli equation \\ \hline
    Nonminimal coupling & $f>0$ & Speed of light &  \\ \hline
    BBMB solution & Ghost appears for $r<2M$ & speed of light & Horizon at $r=M$ \\ \hline
    Models with no scalar hair & $K_X G_4-2 G_{3\phi} G_4+3 G_{4\phi}^2>0$ 
    & speed of light & $G_5$ is irrelevant \\ \hline
    $G^{\mu\nu} \partial_\mu \phi \partial_\nu \phi$ coupling & \eqref{Gkin-stability} & \eqref{Gkin-c1} and \eqref{Gkin-c2}  & Allowed region is shown in Fig.~\ref{fig:allow-region}  \\ \hline
  \end{tabular}
\end{table}

\section{Conclusion}
\label{conclusion}
We have formulated the linear perturbation theory around
static and spherically symmetric spacetime within the framework
of the Horndeski theory, {\em i.e.,} the most general scalar-tensor theory
having second-order field equations both for the metric
and the scalar field.
Following the previous work~\cite{Kobayashi:2012kh}
in which the analysis of the odd-parity perturbations
is presented, we have focused in this paper
on the even-parity perturbations.
Expanding the Horndeski Lagrangian to second order in perturbations
and eliminating the auxiliary variables by use of the constraint equations,
we have derived the reduced Lagrangian that contains only dynamical variables.
The resultant Lagrangian shows that there are two dynamical variables
in the even-parity sector,
with at most first $t$ and $r$ derivatives acting on them, ensuring that the
perturbation equations derived from the Lagrangian
are of second order, as it should be due to
the second-order nature of the Horndeski theory.
We have obtained two conditions for the absence of ghosts:
one coincides with the stability
condition for the odd-parity perturbations and the other provides a new criterion.
The propagation speeds have also been derived. One of them can be
interpreted as the propagation speed of gravitational waves and
is exactly the same as that of the odd-type perturbation.
The other one, the propagation speed of the scalar wave, 
is in general different from that of gravitational waves.
As for the monopole and dipole perturbations, which is absent
in general relativity, there is only one dynamical degree of freedom.
The no-ghost conditions and the propagation speeds
of the monopole and dipole modes are the same as
those of the scalar wave with higher multipoles $\ell \ge 2$.

Our formulation can be applied to any theories belonging to the Horndeski class.
As a demonstration, we have considered several concrete models including
GR, a nonminimally coupled scalar field,
a black hole without scalar hair, the
BBMB solution, and the derivative coupling to the Einstein tensor.
The main results are summarized in Table~\ref{table-1}.

\acknowledgments
This work is supported in part by JSPS Grant-in-Aid for Young
Scientists (B) No.~24740161 (TK),
Grant-in-Aid for Scientific Research on Innovative Areas
No.~25103505 (TS) from The Ministry of Education, Culture, Sports, Science and Technology (MEXT),
and JSPS Postdoctoral Fellowships for Research Abroad (HM).

\appendix

\section{Background equations}
\label{sec-background}

In this Appendix, we summarize the field equations for
a static and spherically symmetric background.
The following field equations were first derived in Ref.~\cite{Kobayashi:2012kh}.

Taking the gauge $C(r)=1$,
the functions $A(r)$ and $B(r)$ in the background metric~(\ref{bgmetric})
can be determined by
solving the field equations supplemented with appropriate boundary conditions.
Substituting the metric~(\ref{bgmetric}) to the action
and varying it with respect to $A$, $B$, $C$ and $\phi$, we obtain
the background field equations as
\begin{eqnarray}
{\cal E}_A=0,
\quad
{\cal E}_B=0,
\quad
{\cal E}_C=0,
\quad
{\cal E}_\phi=\frac{1}{r^2}\sqrt{\frac{B}{A}}\frac{\D}{\D r}\left(r^2\sqrt{AB}\,{\cal J}\right)
-{\cal S}=0, \label{back-eom}
\end{eqnarray}
where
\begin{eqnarray}
{\cal E}_A&:=&K+B\phi'X'G_{3X}-2XG_{3\phi}
+
 \frac{2}{r}\left(\frac{1-B}{r}-B'\right)  G_4
+
 \frac{4B}{r}\left( 
\frac{1}{r}+\frac{X'}{X}+\frac{B'}{B} \right) X G_{4X}
+\frac{8B}{r}XX'G_{4XX}
\nonumber\\&&
-B\phi'\left(\frac{4}{r}+\frac{X'}{X}\right)
G_{4\phi}
+4X G_{4\phi\phi}
 +2B\phi' \left(\frac{4}{r}-\frac{X'}{X} \right) X  G_{4\phi X}
\nonumber\\&&
+
 \frac{B\phi' }{r^2}
\left[(1-3B)\frac{X'}{X}-2B' \right] X G_{5X}
-\frac{2}{r^2} B^2 \phi'XX'G_{5XX}
\nonumber\\&&
-
\frac{2}{r}\left[\frac{1+B}{r}+2B\frac{X'}{X}+B'
\right] X  G_{5\phi}
-\frac{4}{r}B\phi' XG_{5\phi\phi}
 +\frac{4B}{r}\left(\frac{1}{r}-\frac{X'}{X} \right)X^2 
G_{5\phi X},
\\
{\cal E}_B&:=&K-2XK_X
+\left(\frac{4}{r}+\frac{A'}{A}\right)B \phi' XG_{3X}+2X G_{3\phi}
\nonumber\\&&
+\frac{2}{r}\left(\frac{1-B}{r}-B\frac{A'}{A}\right)G_4
-\frac{4}{r}\left(\frac{1-2B}{r}-2 B\frac{A'}{A}\right)XG_{4X}
+\frac{8B}{r}\left(\frac{1}{r}+\frac{A'}{A}\right)X^2G_{4XX}
\nonumber\\&&
-\left(\frac{4}{r}+ \frac{A'}{A}\right)B\phi'G_{4\phi}
-2\left(\frac{4}{r}+ \frac{A'}{A}\right)B\phi'XG_{4\phi X}
+\frac{B\phi'}{r^2}\left(1-5B\right)\frac{A'}{A}XG_{5X}
\nonumber\\&&
-\frac{2B^2\phi'}{r^2}\frac{A'}{A}X^2G_{5XX}
+\frac{2}{r}\left(\frac{1-3B}{r}-3 B\frac{A'}{A}\right)XG_{5\phi}
-\frac{4B}{r}\left(\frac{1}{r}+ \frac{A'}{A}\right)X^2G_{5\phi X},
\\
{\cal E}_C&:=&K+2XG_{3\phi}-X \left( B' \phi'+2B \phi'' \right) G_{3X} 
\nonumber\\&&
-\left[ \frac{1}{r} \sqrt{\frac{B}{A}} \left( r \sqrt{\frac{B}{A}} A' \right)'+\frac{B'}{r} \right]G_4
-
 B\phi' \left( \frac{2}{r} + \frac{A'}{A} + \frac{B'}{B} + 2 \frac{\phi''}{\phi'} \right)
 G_{4\phi}
\nonumber\\&&
+
 B X \left( - \frac{A'^2}{A^2}+\frac{2}{r}\frac{B'}{B}+\frac{A'}{A}
 \frac{B'}{B}+\frac{2(A'+rA'')}{rA} \right) 
 \left( G_{4X}-\frac{1}{2}G_{5\phi} \right)+
 BX' \left( \frac{2}{r}+\frac{A'}{A} \right) \left( G_{4X}-G_{5\phi} \right)+4X G_{4\phi\phi}
\nonumber\\&&
+ 2B\phi' \left( \frac{2}{r} + \frac{A'}{A} - \frac{X'}{X} \right) X  G_{4\phi X}
+ 2r \left( \frac{2}{r}+\frac{A'}{A} \right) XX' G_{4XX}
\nonumber\\&&
-
 \frac{B^2\phi'}{2r} \left[2\frac{A''}{A}-\frac{A'^2}{A^2} + \frac{A'}{A} \left(2\frac{B'}{B}+3\frac{X'}{X} \right) \right] X G_{5X}
-B\phi' X \left( \frac{2}{r}+\frac{A'}{A} \right) G_{5\phi \phi}
\nonumber\\&&
-
\frac{B}{r} \left[ 2\frac{X'}{X}-\frac{rA'}{A} \left( \frac{2}{r}-\frac{X'}{X} \right) \right] X^2
G_{5\phi X}-\frac{B^2 \phi' A'}{rA} XX' G_{5XX},
\\
{\cal J}&:=&\phi'K_X+\left(\frac{4}{r}+\frac{A'}{A}\right)XG_{3X}-2\phi'G_{3\phi}
+2\phi'\left(\frac{1-B}{r^2}-\frac{B}{r}\frac{A'}{A}\right)G_{4X}
-\frac{ 4 B  \phi' }{r}\left(\frac{1}{r}+\frac{A'}{A}\right)XG_{4XX}
\nonumber\\&&
-2\left(\frac{4}{r}+\frac{A'}{A}\right)XG_{4\phi X}
+\frac{1-3B}{r^2}\frac{A'}{A}XG_{5X}-\frac{2B}{r^2}\frac{A'}{A}X^2G_{5XX}
\nonumber\\&&
- 2\phi' \left(\frac{1-B}{r^2}-\frac{B}{r}\frac{A'}{A}\right)G_{5\phi}
+\frac{2B\phi'}{r}\left(\frac{1}{r}+\frac{A'}{A}\right)XG_{5\phi X},
\\
{\cal S}&:=&-K_\phi+2XG_{3\phi\phi}-B\phi'X'G_{3\phi X}
-\left[\frac{B}{2}\left(\frac{A'}{A}\right)^2
+ \frac{2}{r}\left(\frac{1-B}{r}-B'\right)
-\frac{B}{2}\left\{ 2\frac{A''}{A}+\frac{A'}{A}\left(\frac{4}{r}+\frac{B'}{B}\right)
\right\} \right]G_{4\phi}
\nonumber\\&&
+B \left[
\frac{X'}{X}\left(\frac{4}{r}+\frac{A'}{A}\right)
+\frac{4}{r}\left(\frac{1}{r}+\frac{A'}{A}\right)\right] X G_{4\phi X}
+2\left(\frac{1-B}{r^2}-\frac{B}{r}\frac{A'}{A}\right)XG_{5\phi\phi}
\nonumber\\&&
- \frac{B\phi'}{r^2}\left(\frac{X'}{X}+B\frac{A'}{A}\right) X G_{5\phi X},
\end{eqnarray}
and we have set $C(r)=1$ after varying the action.

\section{Expressions of $a_1,~a_2,\cdots$.}
\label{app-a}

The coefficients in the second-order Lagrangian~(\ref{2nd-lag0})
are given explicitly by
\begin{eqnarray}
a_1&=&\sqrt{AB}\,\Xi,\\
a_2&=&\frac{\sqrt{AB}}{2\phi'}\left[
2\phi'\Xi'-\left(2\phi''-\frac{A'}{A}\phi'\right)\Xi
+2r\left(\frac{A'}{A}-\frac{B'}{B}\right){\cal H}
+\frac{2r^2}{B}\left({\cal E}_B-{\cal E}_A\right) \right],\\
a_3&=&-\frac{\sqrt{AB}}{2}\left(\phi'\Xi+2r{\cal H}\right),\\
a_4&=&\sqrt{AB}\,{\cal H},\\
a_5&=&-\sqrt{\frac{A}{B}}r^2\frac{\partial{\cal E}_A}{\partial\phi }=a_2'-a_1'',\\
a_6&=&-\sqrt{\frac{A}{B}} \frac{1}{r\phi'} \left( r {\cal H}'+{\cal H}-{\cal F} \right), \\
a_7&=&a_3'+\frac{r^2}{2} \sqrt{\frac{A}{B}} {\cal E}_B,\\
a_8&=&-\frac{a_4}{2B}, \\
a_9&=&\frac{\sqrt{A}}{r}\frac{\D}{\D r}\left(
r\sqrt{B}{\cal H}
\right)=a_4'+\left( \frac{1}{r}-\frac{A'}{2A} \right) a_4, \\
b_1&=&\frac{1}{2}\sqrt{\frac{B}{A}}{\cal H},
\\
b_2&=&-2\sqrt{\frac{B}{A}}\Xi,
\\
b_3&=&\sqrt{\frac{B}{A}}\frac{1}{\phi'}
\left[
\left(2\phi''+\frac{B'}{B}\phi'\right)\Xi -2r\left(\frac{A'}{A}-\frac{B'}{B}\right){\cal H}
+\frac{2r^2}{B}{\cal E}_A
\right]=\frac{2}{A} \left( a_1'-a_2 \right)+\frac{2r^2}{\sqrt{AB} \phi'} {\cal E}_B,
\\
b_4&=&\sqrt{\frac{B}{A}}\left(\phi'\Xi+2r{\cal H}\right),
\\
b_5&=&-2b_1,\\
c_1&=&-\frac{1}{\sqrt{AB}}\Xi,
\\
c_2&=&-\sqrt{AB} \left( \frac{A'}{2A} \Xi+r\Gamma-\frac{r^2 \phi'}{X} \Sigma \right)
,
\\
c_3&=&r^2\sqrt{\frac{A}{B}} \frac{\partial {\cal E}_B}{\partial \phi},\\
c_4&=&\frac{1}{2}\sqrt{\frac{A}{B}} \Gamma,
\\
c_5&=&
-\frac{1}{2}\sqrt{AB}\left(\phi'\Gamma+\frac{A'}{A}{\cal H}
+\frac{2}{r}{\cal G}\right)
,
\\
c_6&=&\frac{r^2}{2} \sqrt{\frac{A}{B}} \left( \Sigma+\frac{A'B \phi'}{2r^2 A}\Xi+\frac{B\phi'}{r}\Gamma-\frac{1}{2} {\cal E}_B+\frac{B}{r^2} {\cal G}+\frac{A'B}{rA}{\cal H} \right),
\end{eqnarray}

\begin{eqnarray}
d_1&=&b_1,
\\
d_2&=&\sqrt{AB}\,\Gamma,
\\
d_3&=& \frac{\sqrt{AB}}{r^2}\left[
\frac{2r}{\phi'}\left(\frac{A'}{A}-\frac{B'}{B}\right){\cal H}
-r^2\left(\frac{2}{r}-\frac{A'}{A}\right)\frac{\partial{\cal H}}{\partial\phi}
+\frac{2}{B\phi'}\left({\cal F}-{\cal G}\right)\right.
\nonumber\\&&\left.
-\frac{r^2}{2\phi'}\left(2\phi''+\frac{B'}{B}\phi'\right)
\left(\Gamma_1+\frac{2}{r}\Gamma_2\right)
-\frac{2r^2}{B\phi'}({\cal E}_A-{\cal E}_B)
\right]
,
\\
d_4&=&\frac{\sqrt{AB}}{r^2}
\left({\cal G}-r^2{\cal E}_B\right), \\
e_1&=&\frac{1}{2\sqrt{AB}}\left[
\frac{r^2}{X}({\cal E}_A-{\cal E}_B)-\frac{2}{\phi'}\Xi'+\left(\frac{A'}{A}
-\frac{X'}{X}\right)\frac{\Xi}{\phi'}
+\frac{2B}{X}{\cal F}-\frac{2rB}{X}{\cal H}'
-{\cal H}\frac{B^2}{rXA}\frac{\D}{\D r}\left(\frac{r^2A}{B}\right)
\right], \nonumber \\
&=&\frac{1}{A B \phi'} \left[ \left( \frac{A'}{A}+\frac{B'}{2B} \right) a_1 + a_2-2a_1' - 2rB a_6 \right], \\
e_2&=&-\sqrt{AB}\frac{r^2}{X}\Sigma,\\
e_3&=&r^2 \sqrt{\frac{A}{B}} \frac{\partial {\cal E}_\phi}{\partial \phi}, \\
e_4&=&-\frac{1}{4} \sqrt{\frac{A}{B}} {\tilde e_4},
\end{eqnarray}
where $\tilde e_4$ is defined by
\begin{eqnarray} 
\tilde e_4&=&
\frac{2}{X}\left({\cal E}_A-{\cal E}_B\right)
-\frac{2}{\phi'}\Gamma'-\frac{2}{r^2X}\left(1-rB\frac{A'}{A}\right){\cal F}
+\frac{2B}{r^2X}{\cal G}+\frac{2}{r^2X}\left(-2rB\frac{A'}{A}+1-B+rB'\right){\cal H}
\\&&
-\frac{2B}{rX}{\cal G}'-\frac{B}{X}\frac{A'}{A}{\cal H}'
-\frac{2}{r\phi'}\left(2-r\frac{A'}{A}\right)\frac{\partial{\cal H}}{\partial\phi}
+\frac{1}{r^3\phi'}\left(2-r\frac{A'}{A}\right)\left[-2(1-B)+rB\frac{A'}{A}\right]\Xi
\\&&
+\left[
-2\frac{A'}{A}\phi'+\frac{rB}{2}\left(\frac{A'}{A}\right)^2\phi'-\frac{B'}{B}\phi'
+\frac{2}{r}(1-B)\phi'-2\phi''
\right]\frac{\Gamma_1}{\phi'^2}
\\&&
+\left[
-2\frac{A'}{A}\phi'-r(1-B)\left(\frac{A'}{A}\right)^2\phi'-2\frac{B'}{B}\phi'
+\frac{4}{r}(1-B)\phi'-4\phi''
\right]\frac{\Gamma_2}{r\phi'^2}.
\end{eqnarray}

We have the following simple relation among $\Xi,~\Gamma_1$, and $\Gamma_2$:
\begin{equation}
\Xi=\frac{r^2}{2} \Gamma_1+r \Gamma_2-2X G_{5X}.
\end{equation}

\bibliography{draft}
\end{document}